\documentclass{article}
\usepackage{graphicx}
\begin{document}

\title{On the modification of Hamiltonians' spectrum in gravitational quantum mechanics}

\author{Pouria Pedram\thanks{pouria.pedram@gmail.com}\\
{\small Plasma Physics Research Center, Science and Research Campus, Islamic Azad University, Tehran, Iran}}

\date{\today}
\maketitle \baselineskip 24pt

\begin{abstract}
Different candidates of Quantum Gravity such as String Theory,
Doubly Special Relativity, Loop Quantum Gravity and black hole
physics all predict the existence of a minimum observable length or
a maximum observable momentum which modifies the Heisenberg
uncertainty principle. This modified version is usually called the
Generalized (Gravitational) Uncertainty Principle (GUP) and changes
all Hamiltonians in quantum mechanics. In this Letter, we use a
recently proposed GUP which is consistent with String Theory, Doubly
Special Relativity and black hole physics and predicts both a
minimum measurable length and a maximum measurable momentum. This
form of GUP results in two additional terms in any quantum
mechanical Hamiltonian, proportional to $\alpha p^3$ and $\alpha^2
p^4$, respectively, where $\alpha \sim 1/M_{Pl}c$ is the GUP
parameter. By considering both terms as perturbations, we study two
quantum mechanical systems in the framework of the proposed GUP: a
particle in a box and a simple harmonic oscillator. We demonstrate
that, for the general polynomial potentials, the corrections to the
highly excited eigenenergies are proportional to their square
values. We show that this result is exact for the case of a particle
in a box.
\end{abstract}

\textit{Pacs}: {04.60.-m}

\textit{Keywords}: {Quantum gravity; Generalized uncertainty principle; Perturbation analysis.}


\section{Introduction}\label{sec1}
The existence of a minimum measurable length is one of the common
properties of various Quantum Gravity theories such as String
Theory, loop quantum gravity and Doubly Special Relativity.
Moreover, some Gedanken experiments in black hole physics show that
a minimum length of the order of the Planck length arises naturally
from any theory of quantum gravity. Moreover, we can also realizes a
minimal measurable length in the context of spacetime
non-commutativity. We should notice that the minimal observable
length can be probed essentially: we can use a $D0$ brane to probe
the minimal length, but this needs a very long time. In fact, one
can probe the planck length on $D0$ brane if the proposed experiment
lasts an infinite time.

On the other hand, the Heisenberg uncertainty principle does not
exert any restriction on the measurement precision of the particles'
positions or momenta. So, in principle, there is no minimum
measurable length in the usual Heisenberg picture. In the past few
years, many papers have been appeared in the literature to address
the presence of a minimum measurable length by redefinition of the
uncertainty principle in the context of Generalized Uncertainty
Principle (GUP) \cite{1}. This leads to modification of commutation
relations between position and momentum operators in the Hilbert
space. The similar modified commutation relations have also appeared
in Doubly Special Relativity (DSR) theories \cite{2,3}. In fact, DSR
theories also indicate the presence of maximum measurable momenta.

The application of GUP in quantum mechanics would incorporate the
effects of a minimum measurable length on the quantum mechanical
systems which results in the modification of the Hamiltonian (see
\cite{6} and references therein). Since the GUP corrected
Hamiltonian usually contains momentum polynomials of the order of
greater than two, the resulting Schr\"odinger equation has
completely different differential structure. However, since the
effect of the Quantum Gravity is only considerable at the order of
Planck energy, we can use the perturbation method to find the GUP
corrected spectrum of the system. Furthermore, when the corrected
Hamiltonian is naturally perturbative, the perturbation method is
more appropriate than other techniques to find the effect of the
minimum observable length on the energy spectrum.

In this Letter, we consider a recently proposed GUP which is
consistent with String Theory, Doubly Special Relativity and black
hole physics and predicts both a minimum measurable length and a
maximum measurable momentum \cite{main}. First, we obtain the GUP
corrected Hamiltonian to ${\cal{O}}(\alpha^2)$ where $\alpha \sim
1/M_{Pl}c$ is the GUP parameter. Then, using the perturbation
theory, we will obtain the corrected spectrum for two well-known and
instructive cases up to the leading order. These cases consist of a
particle in a box (PB) and  a simple harmonic oscillator (SHO). We
can consider these models as the limiting cases of the potential
$V(x)=|a|x^{2(1+j)}$ where $j=0$ denotes SHO and $j=\infty$ denotes
PB.

\section{A Generalized Uncertainty Principle}\label{sec2}
In a recent paper, Ali \textit{et al} have proposed a generalized
uncertainty principle to address the discreteness of space with the
following commutation relation \cite{main}
\begin{eqnarray}\label{xp}
[x_i, p_j] &=& i \hbar\left[ \delta_{ij}- \alpha \left( p
\delta_{ij} + \frac{p_i p_j}{p} \right) + \alpha^2 \left( p^2
\delta_{ij} + 3p_{i} p_{j} \right) \right],
\end{eqnarray}
where $p^{2} = \sum\limits_{j=1}^{3}p_{j}p_{j} $, $\alpha =
{\alpha_0}/{M_{Pl}c} = {\alpha_0 \ell_{Pl}}/{\hbar}$, $M_{Pl}\equiv$
Planck mass, $\ell_{Pl}\equiv$ Planck length $\approx 10^{-35}m$,
and $M_{Pl} c^2\equiv$ Planck energy $\approx 10^{19}GeV$. Moreover,
the space of positions and momentums is separately assumed
commutative \textit{i.e.} $[x_i,x_j]=[p_i,p_j]=0$. In one-dimension,
above commutation relations result in the following form of the
uncertainty relation to ${\cal{O}}(\alpha^2)$ \cite{5}
\begin{eqnarray}\nonumber
 \Delta x \Delta p &\geq& \frac{\hbar}{2}
\left[ 1 - 2\alpha \langle p\rangle + 4\alpha^2 \langle p^2\rangle
\right]  \\  &\geq& \frac{\hbar}{2} \left[ 1 + \left(
\frac{\alpha}{\sqrt{\langle p^2 \rangle}} +4\alpha^2  \right) \Delta
p^2  +   4\alpha^2 \langle p \rangle^2
 -   2\alpha \sqrt{\langle p^2 \rangle}
\right].
\end{eqnarray}
Note that, the particular form of above inequality implies
\textit{both} a minimum observable length and a maximum observable
momentum at the same time \cite{main}
\begin{eqnarray}
\left\{
\begin{array}{ll}
\Delta x \geq (\Delta x)_{min}  \approx \alpha_0\ell_{Pl},  \\\\
\Delta p \leq (\Delta p)_{max} \approx \frac{M_{Pl}c}{\alpha_0}.
\end{array}
\right.
\end{eqnarray}
Now, let us define
\begin{eqnarray}\label{x0p0}
\left\{
\begin{array}{ll}
x_i = x_{0i},\\\\ p_i = p_{0i} \left( 1 - \alpha p_0 + 2\alpha^2
p_0^2 \right),
\end{array}
\right.
\end{eqnarray}
where $x_{0i}$ and $p_{0i}$ obey the canonical commutation relations
$[x_{0i},p_{0j}]=i\hbar\delta_{ij}$. It is easy to check that using
Eq. (\ref{x0p0}), Eq. (\ref{xp}) is satisfied to
${\cal{O}}(\alpha^2)$. Moreover, from above equation we can
interpret $p_{0i}$ as the momentum operator at low energies
($p_{0i}=-i\hbar \partial/\partial{x_{0i}}$), $p_{i}$ as the
momentum operator at high energies, and $p_0$ as the magnitude of
the $p_{0i}$ vector ($p_{0}^{2}=\sum\limits_{j=1}^{3}p_{0j}p_{0j}$).
It is usually assumed that $\alpha_0$ is of the order of unity. So,
the $\alpha$ dependent terms are important only for high energy
(Planck energy) or high momentum regime. Now, consider the following
general form of the Hamiltonian
\begin{eqnarray}
H=\frac{p^2}{2m} + V(\vec r),
\end{eqnarray}
which by using Eq. (\ref{x0p0}) can be written as
\begin{eqnarray}
H=H_0+\alpha H_1+\alpha^2 H_2+{\cal{O}}(\alpha^3),
\end{eqnarray}
where $H_0=\frac{\displaystyle p_0^2}{\displaystyle2m} + V(\vec r)$
and
\begin{eqnarray}
H_1=-\frac{p_0^3}{m}, \hspace{1cm} H_2=\frac{5p_0^4}{m}.
\end{eqnarray}
In the next two sections, we are interested to study the effect of
$H_1$ and $H_2$ on two one-dimensional quantum mechanical systems
and generalize some results to general polynomial potential cases.

\section{GUP and a particle in a box}\label{sec3}
Here, we apply the GUP formalism to a particle in a box of length
$L$. The boundaries of the box are located at $x = 0$ and $x = L$.
The Hamiltonian of the unperturbed system $H_0=\frac{\displaystyle
p_0^2}{\displaystyle2m}$ results in the following Schr\"odinger
equation
\begin{eqnarray}
H_0\psi_n(x)=E^0_n\psi_n(x),
\end{eqnarray}
where the wave functions should vanish at the boundaries
($\psi_n(0)=\psi_n(L)=0$). So, the corresponding eigenvalues and
normalized eigenfunctions for the unperturbed system are
$E_n^0=\frac{\displaystyle n^2\pi^2 \hbar^2}{\displaystyle2mL^2}$
and $\langle
x|n\rangle\equiv\psi_n(x)=\sqrt{\frac{\displaystyle2}{\displaystyle
L}}\sin(n\pi x/L)$, respectively. Now, using the perturbation
theory, we find the effect of $H_1$ on the energy eigenvalues to
${\cal{O}}(\alpha)$
\begin{eqnarray}
E_n^1=\alpha\langle n|H_1|n\rangle= \frac{2i\alpha
n^3\pi^3\hbar^3}{mL^4} \int_0^L\sin(n\pi x/L)\cos(n\pi x/L)\,d x=0.
\end{eqnarray}
This result shows that, for this case, GUP formalism has no
contribution in energy spectrum to ${\cal{O}}(\alpha)$. To proceed
to the next order, we need to consider both $H_1$ and $H_2$ and
denote their corresponding energy corrections by $E_n^{2,1}$ and
$E_n^{2,2}$, respectively. To obtain $E_n^{2,1}$, we consider the
second order perturbation, namely
\begin{eqnarray}\label{PBE2,1}
E_n^{2,1}=\alpha^2\sum_{k\ne n}^{\infty}\frac{|\langle
k|H_1|n\rangle|^2}{E_n^0-E_k^0}=-\frac{\displaystyle32\alpha^2\pi^2\hbar^4n^6}
 {\displaystyle mL^4}\left\{
\begin{array}{ll}
 \sum_{k=odd}^{\infty}\frac{\displaystyle k^2}{\displaystyle (k^2-n^2)^3}\,\,\,\,\,\, \mbox{for $n$
  even},\\ \\
 \sum_{k=even}^{\infty}\frac{\displaystyle k^2}{\displaystyle (k^2-n^2)^3}\,\,\,\,\,\, \mbox{for $n$
  odd}.
   \end{array}\displaystyle
   \right.
\end{eqnarray}
On the other hand, we have $\sum_{k=odd}^{\infty}\frac{\displaystyle
k^2}{\displaystyle (k^2-n^2)^3}$ for $n$ even
$=\sum_{k=even}^{\infty}\frac{\displaystyle k^2}{\displaystyle
(k^2-n^2)^3}$ for $n$ odd $=\frac{\displaystyle \pi^2}{\displaystyle
64 n^2}$. So, we find
$E_n^{2,1}=-\frac{\displaystyle\alpha^2\pi^4\hbar^4n^4}{\displaystyle
2mL^4}$. Since the contribution of $H_2$ at least is second order in
$\alpha$, after a straightforward calculation, one finds
\begin{eqnarray}\label{PBE2,2}
E_n^{2,2}=\alpha^2\langle n|H_2|n\rangle=\frac{5\alpha^2}{m}\langle
n|p_0^4|n\rangle=\frac{\displaystyle5\alpha^2\pi^4\hbar^4n^4}{\displaystyle
mL^4}.
\end{eqnarray}
Therefore, the total change in energy spectrum is $\Delta
E_n=E_n^{2,1}+E_n^{2,2}=\frac{\displaystyle9\alpha^2\pi^4\hbar^4n^4}{\displaystyle
2mL^4}$ to ${\cal{O}}(\alpha^2)$. Moreover, above equations show
that the contributions of $H_1$ and $H_2$ are in the same order
which result in the modification of the previous results
\cite{nozari}. Note that, we can exactly write $\Delta E_n$ in terms
of unperturbed energy eigenvalues $E_n^0$ as
\begin{eqnarray}\label{delta1}
\Delta E_n=18m\alpha^2{E_n^0}^2,
\end{eqnarray}
or $\frac{\displaystyle \Delta E_n}{\displaystyle E_n^0}\propto
E_n^0$. In other words, the relative change in each energy level is
proportional to its unperturbed energy eigenvalue.

\section{GUP and simple harmonic oscillator}\label{sec4}
Simple harmonic oscillator is one of the most important systems in
quantum mechanics because an arbitrary potential can be approximated
as a harmonic potential at the vicinity of a stable equilibrium
point. The Hamiltonian for this system in the absence of GUP is
given by
\begin{eqnarray}
H_0=\frac{\displaystyle
p_0^2}{\displaystyle2m}+\frac{1}{2}m\omega^2x^2.
\end{eqnarray}
In the quantum mechanical picture ($H_0\psi_n(x)=E_n^0\psi_n(x)$),
this model has well-known eigenvalues and eigenfunctions
\begin{eqnarray}
\psi_{n}(x)&=&\left(\frac{\omega}{\pi}\right)^{1/4}\left[\frac{H_{n}(\sqrt{\omega}
x)}{\sqrt{2^{n}n!}}\right]e^{ -\omega x^{2}/2},\\
E^0_n&=&(n+1/2)\hbar\omega,
\end{eqnarray}
where $H_{n}(x)$ is the Hermite polynomial and the orthonormality
and completeness of the basis functions follow from those of the
Hermite polynomials. We can also express the Hamiltonian in terms of
non-hermitian ladder operators $a=\sqrt{m\omega \over 2\hbar}
\left(x + {i \over m \omega} p_0 \right)$ and $a^{\dagger}= \sqrt{m
\omega \over 2\hbar} \left( x - {i \over m \omega} p_0 \right)$ as
$H_0=\hbar \omega \left(a^{\dagger}a + 1/2\right)$. $a$ and
$a^{\dagger}$  act on an eigenstate of energy $E$ to produce, up to
a multiplicative constant, another eigenstate of energy
$E\pm\hbar\omega$, respectively,
\begin{eqnarray}\label{ladder}
a|n\rangle=\sqrt{n}|n-1\rangle,\hspace{1cm}
a^{\dagger}|n\rangle=\sqrt{n+1}|n+1\rangle,
\end{eqnarray}
with $a|0\rangle=0$. Now, we have enough tools to find the effect of
$H_1$ on energy eigenvalues. Similar to the case of a particle in a
box and without any calculation, we can show that
$E_n^1=\alpha\langle n|H_1|n\rangle$ vanishes also for this case.
Note that, since $p_0$ is proportional to $a$ and $a^{\dagger}$
($p_0=i\sqrt{{\hbar}m\omega \over 2}\left( a^{\dagger}-a \right)$),
$H_1=-\frac{p_0^3}{m}$ consists of odd number of $a$ and
$a^{\dagger}$. Thus, because of Eq. (\ref{ladder}), $E_n^1$ vanishes
for all eigenstates. We can also conclude this result from reality
of energy eigenvalues. To obtain the second order correction of
$H_1$, we need to find the explicit form of $H_1$ in terms of ladder
operators, namely
\begin{eqnarray}
H_1=-\frac{i}{m}\left(\frac{\displaystyle \hbar m
\omega}{\displaystyle
2}\right)^{3/2}({a^{\dagger}}^3-3{a^{\dagger}}^2a-3a^{\dagger}+3a^{\dagger}a^2+3a-a^3),
\end{eqnarray}
where we have used $[a,a^{\dagger}]=1$. So, we have
\begin{eqnarray}\nonumber
\langle k|H_1|n\rangle&=&-\frac{i}{m}\left(\frac{\displaystyle \hbar
m \omega}{\displaystyle
2}\right)^{3/2}\{\sqrt{(n+1)(n+2)(n+3)}\delta_{n+3,k}\\
&-&3(n+1)^{3/2}\delta_{n+1,k}
+3n^{3/2}\delta_{n-1,k}-\sqrt{n(n-1)(n-2)}\delta_{n-3,k}\},
\end{eqnarray}
where $\delta$ is the Kronecker delta symbol. After some algebraic
manipulations, we find the second order correction of $H_1$ as
\begin{eqnarray}\label{sho1}
E_n^{2,1}=\alpha^2\sum_{k\ne n}^{\infty}\frac{|\langle
k|H_1|n\rangle|^2}{E_n^0-E_k^0}=-m\alpha^2\hbar^2\omega^2\left(\frac{\displaystyle
30n^2+30n+11}{\displaystyle 8}\right),
\end{eqnarray}
where we have used $E_n^0-E_k^0=(n-k)\hbar\omega$. Now, let us
consider the contribution of $H_2$ on energy spectrum of SHO. Since
$H_2$ is proportional to the forth power of the momentum, we need to
express $p_0^4$ in terms of ladder operators
\begin{eqnarray}
p_0^4=\left(\frac{m\hbar\omega}{2}\right)^2[12a^{\dagger}a+6{a^{\dagger}}^2a^2+3+({a^{\dagger}}^4-4{a^{\dagger}}^3a-6{a^{\dagger}}^2-4a^{\dagger}a^3-6a^2+a^4)].
\end{eqnarray}
Note that, in above equation, the terms which are in parentheses
have unequal number of $a$ and $a^{\dagger}$ and consequently they
do not contribute in $\langle n|H_2|n\rangle$. So, we can write the
leading order contribution of $H_2$ as
\begin{eqnarray}\label{sho2}
E_n^{2,2}&=&\alpha^2\langle
n|H_2|n\rangle=\frac{5\alpha^2}{m}\left(\frac{m\hbar\omega}{2}\right)^2\langle
n|12a^{\dagger}a+6{a^{\dagger}}^2a^2+3|n\rangle=m\alpha^2\hbar^2
\omega^2\left(\frac{\displaystyle30n^2+30n+15}{\displaystyle
4}\right).
\end{eqnarray}
Therefore, using Eq. (\ref{sho1}) and Eq. (\ref{sho2}), the total
effect of GUP on energy levels to ${\cal{O}}(\alpha^2)$ is $\Delta
E_n=E_n^{2,1}+E_n^{2,2}=\frac{\displaystyle 1}{\displaystyle
8}m\alpha^2\hbar^2\omega^2\left(30n^2+30n+19\right)$. Moreover, Eqs.
(\ref{sho1}) and (\ref{sho2}) show that, also in this case, the
corrections of $H_1$ and $H_2$ to energy levels are in the same
order which modify the previous results for SHO \cite{5}. For large
values of $n$ ($n\gg1$), we have $\Delta E_n\cong\frac{\displaystyle
15 }{\displaystyle 4}m\alpha^2\hbar^2\omega^2 n^2$ and $E_n^0\cong
n\hbar\omega$. Therefore, for large $n$, we can write
\begin{eqnarray}\label{delta2}
\Delta E_n\cong\frac{\displaystyle 15}{\displaystyle 4}m\alpha^2
{E_n^0}^2,
\end{eqnarray}
or $\frac{\displaystyle \Delta E_n}{\displaystyle E_n^0}\propto
E_n^0$. So, also in this case, the relative change in each energy
level is proportional to its energy eigenvalue.

Since now, we have studied two limiting cases of
$V(x)=|a|x^{2(j+1)}$ potentials ($j\in$ positive integers), where
$j=0,\infty$ correspond to SHO and PB, respectively. Moreover, we
found that the relation $\frac{\displaystyle \Delta
E_n}{\displaystyle E_n^0}\propto E_n^0$ is exact for all energy
levels of PB (\ref{delta1}) and is valid for high energy levels of
SHO (\ref{delta2}). Since these two systems are limiting cases, we
can generalize this result for all other values of $0<j<\infty$.

To justify this generalization we can use a general operational
procedure called the Factorization Method \cite{Infeld}. In this
method, the Hamiltonian of the system is written as the
multiplication of two ladder operators plus a constant
($H=a^{\dag}a+E$). Then, these operators are used to obtain the
Hamiltonian's eigenfunctions. In general, in contrast to the case of
SHO, one ladder operator is not enough to form all the Hamiltonian's
eigenfunctions and for each eigenfunction a ladder operator is
needed.

The procedure of finding the ladder operators and the eigenfunctions
consists of some steps; We find operators $a_{1}, a_{2}, a_{3},
\ldots$ and real constants $E_{1}^0, E_{2}^0, E_{3}^0, \ldots$ from
the following recursion relations
\begin{equation}\label{factorization1}
      \begin{array}{l}
            a^{\dag}_{1}a_{1}+E_{1}^0=H_0,\\ \\
            a^{\dag}_{2}a_{2}+E_{2}^0=a_{1}a^{\dag}_{1}+E_{1}^0,\\
            \\
            a^{\dag}_{3}a_{3}+E_{3}^0=a_{2}a^{\dag}_{2}+E_{2}^0, \ldots,
      \end{array}
\end{equation}
or generally
\begin{equation}\label{factorization2}
a^{\dag}_{n+1}a_{n+1}+E_{n+1}^0=a_{n}a^{\dag}_{n}+E_{n}^0,
\hspace{1cm}n=1,2,\ldots,
\end{equation}
where the real constants $E_{n}^0$ are the eigenvalues of the
Hamiltonian and the operators $a_{n},a^{\dagger}_n$ are the ladder
operators used to form the eigenfunctions. Also, assume that there
exists a null eigenfunction (root function) $|\xi_{n}\rangle$ with
zero eigenvalue for each $a_{n}$, namely
\begin{equation}\label{eigenvector1}
a_{n}|\xi_{n}\rangle=0.
\end{equation}
Hence, $E_{n}^0$ is the $n^{th}$ eigenvalue of the Hamiltonian with
the following corresponding eigenfunction
\begin{equation}\label{eigenvector2}
|n\rangle=a^{\dag}_{1}a^{\dag}_{2}...a^{\dag}_{n-1}|\xi_{n}\rangle.
\end{equation}
Because of recursion relations each of the annihilation operators
($a_{n}$) should contain a linear momentum term. Thus, $a_{n}$ can
be written as
\begin{equation}\label{fj1}
a_{n}=\frac{1}{\sqrt{2m}}(p_0+i f_{n}(x)),
\end{equation}
where $f_{n}(x)$ is a real function of $x$ ($f_n(x)=-m\omega x$ for
SHO and $f_n(x)=\frac{\displaystyle n\pi\hbar}{\displaystyle L}
\cot\left(\frac{\displaystyle\pi x}{\displaystyle L}\right)$ for
PB). From above equation, we have
$p_0=\sqrt{2m}(a_1+a_1^{\dagger})$. So, the relevant terms of
$p_0^4$ which have non-zero contribution in the expectation value of
$H_2=\frac{\displaystyle5p_0^4}{\displaystyle m}$ are
\begin{eqnarray}\label{p04}
(p_0^4)^{\mbox{\footnotesize{relevant}}}=4m^2\left[(a_1^{\dagger}a_1)^2+(a_1a_1^{\dagger})^2+
a_1^2{a_1^{\dagger}}^2+{a_1^{\dagger}}^2a_1^2+a_1{a_1^{\dagger}}^2a_1+a_1^{\dagger}a_1^2a_1^{\dagger}\right],
\end{eqnarray}
where $[a_1,a_1^{\dagger}]=-\frac{\displaystyle\hbar}{\displaystyle
m}\frac{\displaystyle df_1(x)}{\displaystyle dx}$. Note that, for
high energy levels, we have $\langle
n|a_1a_1^{\dagger}|n\rangle\cong \langle
n|a_1^{\dagger}a_1|n\rangle\cong \langle n|H_0|n\rangle$. In other
words, in this limit, the effect of
$\frac{\displaystyle\hbar}{\displaystyle m}\frac{\displaystyle
df_1(x)}{\displaystyle dx}$ is negligible with respect to the
Hamiltonian $H_0$. Now, let us verify this result for two studied
cases. For PB we have $\frac{\displaystyle\hbar}{\displaystyle
m}\frac{\displaystyle df_1(x)}{\displaystyle dx}=\frac{\displaystyle
-2E_1^0}{\displaystyle \sin^2(\pi x/L)}$ which results in
$a_1a_1^{\dagger}=a_1^{\dagger}a_1-\frac{\displaystyle\hbar}{\displaystyle
m}\frac{\displaystyle df_1(x)}{\displaystyle
dx}=H_0-E_1^0+\frac{\displaystyle2E_1^0}{\displaystyle\sin^2(\pi
x/L)}$. So, for $n\gg1$ one finds
\begin{eqnarray}
\langle n|a_1a_1^{\dagger}|n\rangle&=&\langle
n|H_0|n\rangle-E_1^0+2E_1^0(2n)\cong\frac{n^2\pi^2\hbar^2}{2mL^2}\cong
\langle n|a_1^{\dagger}a_1|n\rangle.
\end{eqnarray}
For the case of simple harmonic oscillator, we have
$\frac{\displaystyle\hbar}{\displaystyle m}\frac{\displaystyle
df_1(x)}{\displaystyle dx}=-\hbar\omega$ and
$a_1a_1^{\dagger}=H_0+\frac{1}{2}\hbar\omega$ which result in
\begin{eqnarray}
\langle n|a_1a_1^{\dagger}|n\rangle=\langle
n|H_0|n\rangle+\frac{1}{2}\hbar\omega\cong n\hbar\omega\cong \langle
n|a_1^{\dagger}a_1|n\rangle.
\end{eqnarray}
Therefore, in high energy regime, to calculate the expectation value
of $H_2$ all terms in Eq.~(\ref{p04}) act as the first term. So, we
have
\begin{eqnarray}\label{factor}
\alpha^2\langle n|H_2|n\rangle\cong{\cal{O}}(1)m\alpha^2\langle
n|H_0^2|n\rangle\cong{\cal{O}}(1)m\alpha^2{E_n^0}^2.
\end{eqnarray}
Moreover, because of the normalization condition of the energy
eigenstates (\ref{eigenvector2}), for $n\gg1$ we have
\begin{eqnarray}\label{high}
a_1|n\rangle\sim\sqrt{E_n^0}|n-1\rangle,\hspace{1cm}a_1^{\dagger}|n\rangle\sim\sqrt{E_n^0}|n+1\rangle,
\end{eqnarray}
which also imply Eq.~(\ref{factor}) in a more straightforward
manner. On the other hand, the hermiticity of the Hamiltonian shows
that, for a general polynomial potential, $H_1$ has no first order
contribution in the energy spectrum. Also, the explicit form of
$H_1=-\frac{\displaystyle p_0^3}{\displaystyle m}$ in terms of
ladder operators ($a_1,a_1^{\dagger}$) and Eq.~(\ref{high}) show
that $\langle k|H_1|n\rangle\sim {E_n^0}^{3/2}$ and consequently the
second order perturbation correction $E_n^{2,1}$ is also
proportional to ${E_n^0}^2$.

On the other hand, for $n\gg1$ the spectrum of
$H_0=\frac{\displaystyle p_0^2}{\displaystyle2m}+|a|x^{2(j+1)}$
coincides with
\begin{eqnarray}\label{hamilton}
H_0=\frac{\displaystyle
p_0^2}{\displaystyle2m}+|a|x^{2(j+1)}+b\,x^{2(j+1)-1}+c\,x^{2(j+1)-2}+\ldots\,,
\end{eqnarray}
which can be obtained from Sommerfeld-Wilson quantization rule
\begin{eqnarray}\label{sommer}
\oint p_x\,d x=nh,\hspace{3cm}n=1,2,\ldots,
\end{eqnarray}
in high energy limit, where $p_x=\sqrt{2m(E-V(x))}$. In this energy
limit, for the same value of the classical turning points
($x_{TP}\gg1$), this integral for $V_1(x)=|a|x^{2(j+1)}$ is
approximately equal to
$V_2(x)=|a|x^{2(j+1)}+b\,x^{2(j+1)-1}+c\,x^{2(j+1)-2}+\ldots\,$.
Because, the dominance of $V_2$ over $V_1$ is around $x\approx0$
which for $E\gg1$ (hereafter we choose $\frac{\hbar^2}{2m}=1$) does
not alter considerably the value of $E-V$ in the integrand.
Moreover, since $|x_{TP}|\equiv  L\gg1$, we have
$E_{1,n}=V_1(L)\cong V_2(L)=E_{2,n}$. For instance, for
$V_1(x)=x^4$, using Eq.~(\ref{sommer}), we have $E_{1,n}=L^4=\beta
n^{4/3}$ where
$\beta=\left(\frac{\sqrt{\pi}\Gamma(7/4)}{\Gamma(5/4)}\right)^{4/3}$.
For $V_2(x)=x^4+x^2$ we have $ \int_{-L}^{L}
\sqrt{L^4+L^2-x^4-x^2}\,d x=\int_{-L}^{L} \sqrt{L^4-x^4}\,d
x+{\cal{O}}(1)L=n\pi$ which results in $E_{2,n}= L^4+L^2\cong\beta
n^{4/3}+{\cal{O}}(1)n^{2/3}$. So, for $n\gg1$ we have $E_{1,n}\cong
E_{2,n}$.

Now, following Eq. (\ref{delta1}) and Eq. (\ref{delta2}), for the
general form of the Hamiltonian (\ref{hamilton}), we can write the
following relation
\begin{eqnarray}
\Delta E_n\cong {\cal{O}}(1)m\alpha^2 {E_n^0}^2,
\end{eqnarray}
which is an approximate relation for $n\gg1$.

\section{Conclusions}\label{sec5}
In this Letter, we have considered the consequence of a Generalized
(Gravitational) Uncertainty Principle on the spectrum of some
quantum mechanical systems. This principle comes from the presence
of a minimum observable length and modifies all Hamiltonians in
quantum mechanics. Following the recently proposed GUP which is
consistent with String Theory, Doubly Special Relativity, black hole
physics and also implies a maximum observable momentum, we found the
energy eigenvalues of a particle in a box and a simple harmonic
oscillator up to the second order of the minimum length
($\ell_{Pl}$). We showed that, for the case of a particle in a box,
the corrections to the eigenenergies are exactly proportional to
their square values. We also concluded that, for the general
polynomial potentials in the form
$V(x)=|a|x^{2(j+1)}+bx^{2(j+1)-1}+cx^{2(j+1)-2}+...\,$, this result
is approximately valid for highly excited eigenenergies.

\section*{Acknowledgments}
I would like to thank K. Nozari for useful discussion and comments.

\end{document}